\documentclass[conference,10pt]{IEEEtran}
\IEEEoverridecommandlockouts
\usepackage{setspace}               % set line space
\usepackage{stfloats}
\usepackage{cite}
\usepackage{amsmath,amssymb,amsfonts}
\usepackage{upgreek}
\usepackage{graphicx}
\usepackage{subfigure}
\usepackage{textcomp}
\usepackage{xcolor}
\usepackage{algorithm}
\usepackage[noend]{algpseudocode}
\usepackage{amsmath}
\usepackage[caption=false,font=normalsize,labelfont=sf,textfont=sf]{subfig}
\usepackage[linewidth=1pt]{mdframed}
\usepackage[mathscr]{eucal}
\usepackage{epstopdf}
\usepackage{cases}
\usepackage{xcolor}
\usepackage{bbding}
\usepackage{cleveref}
\usepackage{multicol}       % table
\usepackage{multirow}       % table
\usepackage{array}          % table
\usepackage{makecell}
\usepackage{flushend}
\usepackage{booktabs}
\usepackage{geometry}
\definecolor{crimson}{RGB}{192,0,0}         % color crimson
\definecolor{navy}{RGB}{47,85,151}         % color crimson
\usepackage{colortbl}

\def\BibTeX{{\rm B\kern-.05em{\sc i\kern-.025em b}\kern-.08em
    T\kern-.1667em\lower.7ex\hbox{E}\kern-.125emX}}

\begin{document}
%----------------------------title&author&thanks----------------------------

\title{Distributed Collaborative User Positioning for Cell-Free Massive MIMO with Multi-Agent Reinforcement Learning}
\author{{Ziheng~Liu,~\IEEEmembership{Student Member,~IEEE}, Jiayi~Zhang,~\IEEEmembership{Senior Member,~IEEE}, Enyu~Shi,~\IEEEmembership{Student Member,~IEEE},}\\ {Yiyang~Zhu, Derrick~Wing~Kwan~Ng,~\IEEEmembership{Fellow,~IEEE}, and Bo~Ai,~\IEEEmembership{Fellow,~IEEE}}
\thanks{This work was supported by ZTE Industry-University-Institute Cooperation Funds under Grant No. IA20240709018.}
\thanks{Z. Liu, J. Zhang, E. Shi, Y. Zhu, and B. Ai are with the School of Electronic and Information Engineering and also with the Frontiers Science Center for Smart High-Speed Railway System, Beijing Jiaotong University, Beijing 100044, China (e-mail: \{23111013, zhangjiayi, 21111047, 21251058, boai\}@bjtu.edu.cn).}
\thanks{D. W. K. Ng is with the School of Electrical Engineering and Telecommunications, University of New South Wales, NSW 2052, Australia (e-mail: w.k.ng@unsw.edu.au).}}
\addtolength{\topmargin}{0.005cm}
\maketitle
%----------------------------abstract----------------------------
\begin{abstract}
In this paper, we investigate a cell-free massive multiple-input multiple-output system, which exhibits great potential in enhancing the capabilities of next-generation mobile communication networks. We first study the distributed positioning problem to lay the groundwork for solving resource allocation and interference management issues. Instead of relying on computationally and spatially complex fingerprint positioning methods, we propose a novel two-stage distributed collaborative positioning architecture with multi-agent reinforcement learning (MARL) network, consisting of a received signal strength-based preliminary positioning network and an angle of arrival-based auxiliary correction network. Our experimental results demonstrate that the two-stage distributed collaborative user positioning architecture can outperform conventional fingerprint positioning methods in terms of positioning accuracy.
\end{abstract}
%----------------------------keywords----------------------------

%\newpage
\IEEEpeerreviewmaketitle
\section{Introduction}
Cell-free massive multiple-input multiple-output (CF mMIMO), a promising technology for future wireless communication systems \cite{[1],[40],[41]}, has gained significant attention due to its potential in achieving uniform spectral efficiency (SE). Compared with conventional cellular mMIMO technologies \cite{[21]}, cell-free mMIMO presents an advanced network architecture that utilizes a large number of collaborative access points (APs) \cite{[22],[23],[43]}. Consequently, all APs coherently serve all user equipments (UEs) without any cell boundaries to achieve spatial multiplexing with shared time-frequency resources in cell-free mMIMO systems \cite{[3]}. Moreover, cell-free mMIMO is also a significant leap in conventional mMIMO technologies in alleviating inter-cell interference \cite{[44]}, effectively eliminating the basic limitations of inter-cell interference on the performance of dense cellular systems.

Recently, with the widespread application of various positioning technologies in mobile communication networks, it has greatly aroused the interest of academia in developing precise positioning schemes \cite{[28],[11],[8]}. Specifically, the strategy of utilizing the rich information provided by multipath wireless channels to achieve user positioning has gained considerable momentum, such as geometry positioning \cite{[28]}, fingerprint positioning \cite{[11]}, and machine learning-based positioning \cite{[8]}. In particular, the most commonly adopted positioning information among these schemes are received signal strength (RSS) and angle of arrival (AOA), where the former mainly maps the user's distance information, while the latter mainly maps the corresponding angle information. However, for conventional schemes that include fingerprint and geometric, their high computational complexity makes them unsuitable for practical large-scale networks. On the other hand, novel model-free machine learning schemes mostly tend to adopt supervised learning architectures, which is unrealistic because obtaining prior optimal output data is challenging.

In contrast, multi-agent reinforcement learning (MARL) has emerged as a disruptive technique in various research domains due to its outstanding scalability and effectiveness \cite{[29],[42]}. Specifically, MARL is a distributed extension based on the existing single-agent RL architecture, where multiple agents interact with each other to jointly complete complex tasks. Based on the above advantages, MARL has also been extensively investigated for solving various wireless resource allocation problems in cell-free mMIMO systems \cite{[31],[5]}. For example, the authors in \cite{[5]} investigated an innovative double-layer power control architecture with MARL network, providing a dynamic strategy for solving high-dimensional signal processing problems while effective balancing computational complexity and system performance.
Unfortunately, due to the high demand for positioning accuracy, it is challenging to simply adopt MARL to achieve the set goals in actual cell-free mMIMO systems.
Hence, there is a growing necessity to develop specialized MARL networks that cater to specific types of observed states, enabling collaborative optimization of various problems.

Motivated by the above observations, we investigate a cell-free mMIMO system, and derive RSS values and the angular domain channel power matrix for positioning model. Furthermore, we introduce a distributed collaborative positioning network with MARL to optimize the user positioning problem, which primarily relies on RSS information for initial positioning, supplemented by AOA information for positioning correction, thereby enhancing positioning accuracy.
\addtolength{\topmargin}{0.015cm}
\section{System Model}
\begin{figure*}[t]
\centering
    \includegraphics[scale=0.6]{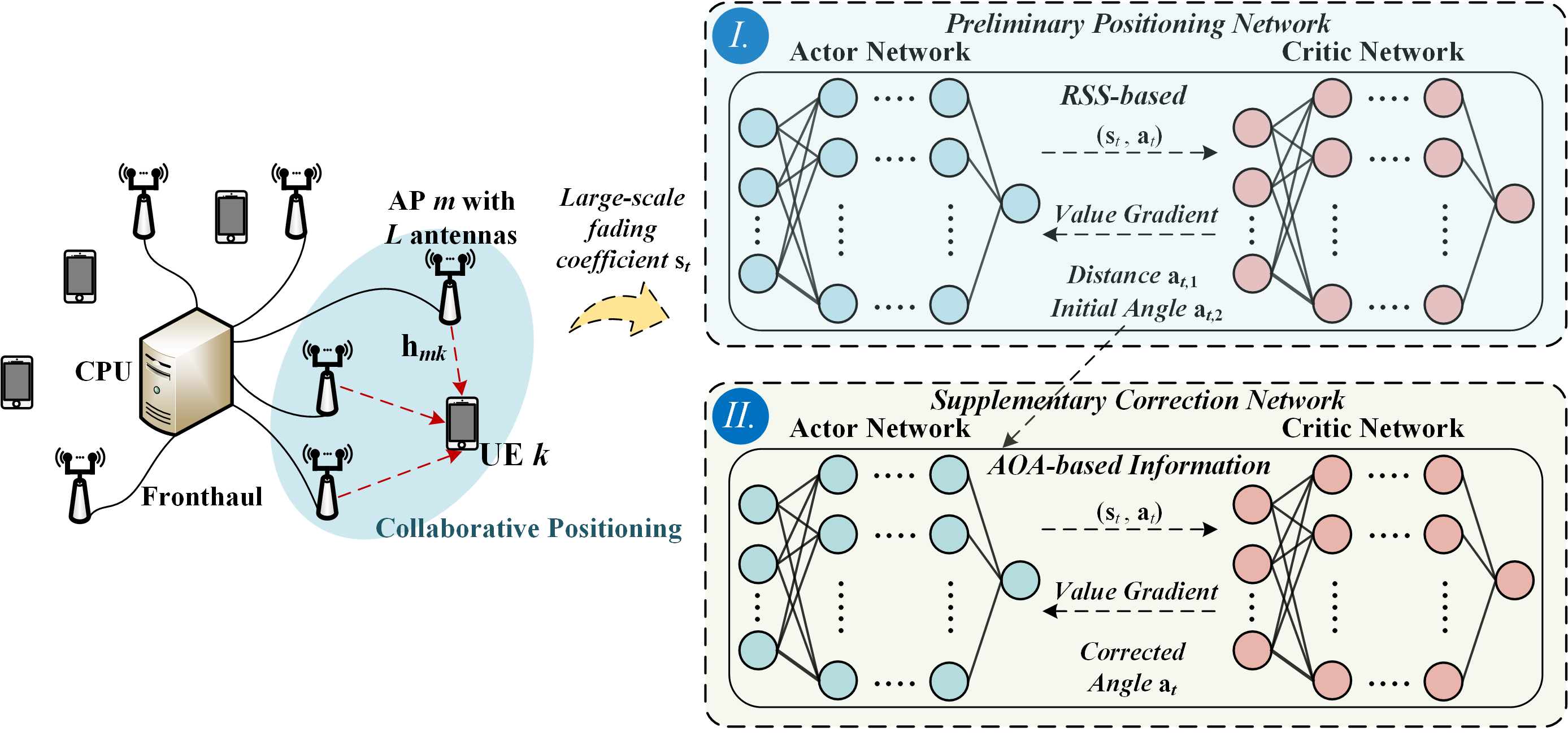}
    \caption{Illustration of the global positioning architecture mainly consists of three parts: a cell-free mMIMO system with collaborative positioning mechanism, a RSS-based preliminary positioning network, and an AOA-based auxiliary correction network.
    \label{fig1}}
\end{figure*}
%\addtolength{\topmargin}{0.13cm}
In this article, we investigate a hardware-constrained cell-free mMIMO system consisting of $M$ APs and $K$ UEs arbitrarily distributed in a large coverage area and denote $\mathcal{M}=\{1,\ldots,M\}$ and $\mathcal{K}=\{1,\ldots,K\}$. Each AP is equipped with $L$ antennas and directly connected to the central processing unit (CPU) via fronthaul links, as illustrated in Fig. 1.
\subsection{Channel Model}
In cell-free mMIMO systems, we consider the channel based on the quasi-static block fading channel model, where the channel is frequency flat and static in each coherent time block, which is described as \cite{[8]}
\begin{equation}
\setcounter{equation}{1}
\mathbf{h}_{mk} = \sqrt{\frac{1}{N_{mk}}}\sum_{n=1}^{N_{mk}}\sqrt{\beta_{mk}}\alpha_{mk}^{n}\mathbf{a}(\theta_{mk}^{n}),
\label{eq1}
\end{equation}
where $N_{mk}$ is the number of scattering paths, $\beta_{mk}$ represents large-scale fading coefficients and $\alpha_{mk}^{n} \sim {{\mathcal N}_\mathbb{C}}(0,1)$ denotes small-scale Rayleigh fading coefficients of the $n$-th scattering path, $n \in \{1,\dots,N_{mk}\}$. In addition, $\mathbf{a}(\theta_{mk}^{n})=[1, e^{-j2\pi\frac{\Delta}{\lambda}\mathrm{cos}(\theta_{mk}^{n})}, \ldots, e^{-j2\pi\frac{(L-1)\Delta}{\lambda}\mathrm{cos}(\theta_{mk}^{n})}]^{T} \in \mathbb{C}^{L \times 1}$ is the array steering vector, where $\lambda$, $\Delta$, and $\theta_{mk}^{n}$ are the wavelength, antenna spacing, and AOA, respectively.
\subsection{Channel Estimation}
In this phase, to improve the accuracy of the extracted positioning information, we assign $\tau_p$ mutually orthogonal pilot sequences $\boldsymbol{\phi}_{1}, \ldots, \boldsymbol{\phi}_{\tau_p}$ to all UEs, satisfying $K = \tau_p$ and $\|\boldsymbol{\phi}_{k}\|^2=\tau_{p}$. Therefore, the received signal $\mathbf{Y}_m^p \in \mathbb{C}^{L \times \tau_p}$ at AP $m$ can be described as
\begin{equation}
\setcounter{equation}{2}
\mathbf{Y}_m^p = \sum_{i \in \mathcal{K}}\sqrt{p_i}\mathbf{h}_{mi}\boldsymbol{\phi}_{i}^{T} + \mathbf{N}_m,
\label{eq3}
\end{equation}
where $p_i$ and $\mathbf{N}_m \in \mathbb{C}^{L \times \tau_p}$ are the transmit power of UE $i$ and the noise with independent ${{\mathcal N}_\mathbb{C}}(0,\sigma^2)$ entries and the noise power $\sigma^2$, respectively. Then, all APs perform coherent linear processing on the received pilot signal $\mathbf{Y}_m^p$, $m \in \mathcal{M}$. Correspondingly, the received pilot signal of UE $k$ at AP $m$ can multiplied by the conjugate pilot sequence $\boldsymbol{\phi}_{k}^\ast$ to obtain, which can be expressed as
\begin{equation}
\begin{split}
\setcounter{equation}{3}
\mathbf{y}_{mk}^p \triangleq {\frac{1}{\sqrt{\tau_p}}}\mathbf{Y}_m^p\boldsymbol{\phi}_{k}^\ast =\sqrt{p_k\tau_p}\mathbf{h}_{mk} + \mathbf{n}_{mk},
\end{split}
\end{equation}
where $\mathbf{n}_{mk} \triangleq \mathbf{N}_m\boldsymbol{\phi}_{k}^\ast/\sqrt{\tau_p} \sim {{\mathcal N}_\mathbb{C}}(\mathbf{0},\sigma^2\mathbf{I}_L)$ is the resulting noise. We can notice that equation (4) is applicable to any linear estimator. Here, we consider adopting a least square (LS) estimator with lower computational complexity \cite{[38]} to minimize $\|\mathbf{y}_{mk}^p - \sqrt{p_k\tau_p}\mathbf{h}_{mk}\|$. Then, the estimated channel can be expressed as
${\mathbf{\hat{h}}}_{mk} =\frac{1}{\sqrt{p_k\tau_p}}(\sqrt{p_k\tau_p}\mathbf{h}_{mk} + \mathbf{n}_{mk})$.
\subsection{Positioning Information Extraction}
To achieve precise user positioning, the goal of each AP is to determine positioning information from the obtained signals and obtain the position of each UE through mutual assistance. In the following subsections, we discuss two different methods for extracting positioning information.
\subsubsection{RSS Fingerprint Extraction}
After all APs receive pilot signals from UE $k$, we can obtain the corresponding RSS value $\mathbf{\Psi}_{k}=[\psi_{1k},\ldots,\psi_{Mk}]^\mathrm{T}$ with $\psi_{mk} = p_k\tau_p\|\mathbf{h}_{mk}\|^2$, which is determined solely by channel information $\mathbf{h}_{mk}$.
However, the existence of small-scale fading may lead to random fluctuations in RSS value $\psi_{mk}$, we can adopt channel hardening techniques to suppress this interference, so that the normalized instantaneous channel gain asymptotically converges to the deterministic average channel gain \cite{[36]}. Therefore, the RSS value $\psi_{mk}$ can be further simplified as
\begin{equation}
\setcounter{equation}{4}
\psi_{mk} \approx p_k\tau_p \mathbb{E}\Big\{\|\mathbf{h}_{mk}\|^2\Big\} = Lp_k\tau_p\beta_{mk}.
\label{eq4}
\end{equation}

\addtolength{\topmargin}{-0.02cm}
Obviously, note that equation (4) indicates the RSS value $\psi_{mk}$ is directly proportional to the large-scale fading coefficient and the number of antennas.
\subsubsection{AOA Fingerprint Extraction}
Considering the RSS value $\psi_{mk}$ only contains rough channel information, making it impossible to achieve high-precision user positioning in complex communication environments. In contrast, AOA values composed of large-scale statistical channel information are a better alternative solution, which is related to the spatial sparse structure of the channel \cite{[10]}. Therefore, we can adopt the DFT matrix $\mathbf{F} \in \mathbb{C}^{L \times L}$ with $[\mathbf{F}]_{i,j}=e^{-j2\pi\frac{(i-1)(j-1)}{L}}$ to map the estimated channel $\mathbf{\hat{h}}_{mk}$ into the angular domain channel $\mathbf{g}_{mk}$, which can be expressed as $\mathbf{g}_{mk} = \mathbf{F}\mathbf{\hat{h}}_{mk}$. Correspondingly, the angular domain channel response matrix $\mathbf{G}_k \in \mathbb{C}^{L \times M}$ can be given by
\begin{equation}
\setcounter{equation}{5}
\mathbf{G}_k=\Big[\mathbf{g}_{1k},\ldots,\mathbf{g}_{Mk}\Big]=\Big[\mathbf{F}\mathbf{\hat{h}}_{1k},\ldots,\mathbf{F}\mathbf{\hat{h}}_{Mk}\Big].
\label{eq5}
\end{equation}

Moreover, to further suppress the impact of channel fluctuations caused by small-scale fading, we adopt statistical channel data as a measure \cite{[4]}, and the angular domain channel power matrix can be expressed as $\mathbf{\Theta}_k \triangleq \mathbb{E}\{\mathbf{G}_k \odot \mathbf{G}_k^\ast\} \in \mathbb{R}^{L \times M}$,
where $[\mathbf{\Theta}_k]_{l,m} = \mathbb{E}\{|[\mathbf{G}_k]_{n,m}|^2\}$. Note that $\mathbf{\Theta}_k$ provides an effective description of AOA and channel power distribution, which helps improve positioning accuracy.
\section{Distributed Cooperative User Positioning}
In this section, we first study the user positioning model. Then, considering the situation where the actual positions of all UEs is unknown, the positioning performance is measured by introducing a similarity model.
\subsection{Positioning Model}
To quantitatively analyze the positioning accuracy in studied cell-free mMIMO systems, we adopt the root mean square error (RMSE) to evaluate the overall positioning error, which can be described as
\begin{equation}
\setcounter{equation}{6}
e_\text{RMSE}=\sqrt{\frac{1}{K}\sum_{k \in \mathcal{K}}\Big((\hat{x}_{k}-x_k)^2+(\hat{y}_{k}-y_k)^2\Big)},
\label{eq7}
\end{equation}
where $(x_k,y_k)$ and $(\hat{x}_{k},\hat{y}_{k})$ are the actual position and estimated position of UE $k$, respectively. However, due to the fact that the actual position of each UE is unknown during the implementation of user positioning, equation (6) cannot be solved. By contrast, considering that the corresponding channel state information is known in advance, we can replace the original positioning model $e_\text{RMSE}$ with the correlation between the estimated position and the actual position to achieve user positioning.
\subsection{Similarity Model}
In this section, we can adopt AOA $\mathbf{\Theta}_k$ or RSS $\mathbf{\Psi}_{k}$ to establish positioning similarity between actual and estimated positions. In theory, the larger the positioning similarity coefficient, the higher the correlation between the actual position and the estimated position. However, the obtained similarity coefficients may be damaged by the wireless transmission environment and cannot accurately reflect the correlation in physical position. Therefore, we can adopt the joint similarity criterion to improve user positioning by supplementing the existing RSS value with AOA value, which can be given by
\begin{equation}
\setcounter{equation}{7}
\mathbf{\Xi}(\mathbf{\Theta}_k,\mathbf{\Psi}_k,\hat{\mathbf{\Theta}}_{k},\hat{\mathbf{\Psi}}_{k})=\frac{\mathbf{\Xi}^{{a}}(\mathbf{\Theta}_k,\hat{\mathbf{\Theta}}_{k})}{\mathbf{\bar{\Xi}}^{{r}}(\mathbf{\Psi}_k,\hat{\mathbf{\Psi}}_{k})},
\label{eq8}
\end{equation}
where $\mathbf{\Xi}(\mathbf{\Theta}_k,\mathbf{\Psi}_k,\hat{\mathbf{\Theta}}_{k},\hat{\mathbf{\Psi}}_{k})$ quantifies the joint correlation between AOA and RSS values.
$\mathbf{\Xi}^{{a}}_m(\mathbf{\Theta}_k,\hat{\mathbf{\Theta}}_{k}) \in [0,1]$ and $\mathbf{\bar{\Xi}}^{{r}}(\mathbf{\Psi}_k,\hat{\mathbf{\Psi}}_{k}) \in [0,1]$ are the AOA-based and the normalized RSS-based similarity coefficient, respectively, which can be expressed as
\begin{equation}
\begin{split}
\setcounter{equation}{8}
\mathbf{\Xi}^{{a}}(\mathbf{\Theta}_k,\hat{\mathbf{\Theta}}_{k})=\frac{1}{\sqrt{M}}\sum_{m \in \mathcal{M}} \frac{[\mathbf{\Theta}_k]_{:,m}^{T}[\hat{\mathbf{\Theta}}_{k}]_{:,m}}{\|[\mathbf{\Theta}_k]_{:,m}\|\|[\hat{\mathbf{\Theta}}_{k}]_{:,m}\|},
\end{split}
\end{equation}
and
\begin{equation}
\begin{split}
\setcounter{equation}{9}
\mathbf{\bar{\Xi}}^{{r}}(\mathbf{\Psi}_k,\hat{\mathbf{\Psi}}_{k})=\frac{\sqrt{\sum_{m \in \mathcal{M}}|\psi_{mk}-\hat{\psi}_{mk}|^2}}{\max_{\forall i \in \mathcal{K}}\Big\{\sqrt{\sum_{m \in \mathcal{M}}|\psi_{mi}-\hat{\psi}_{mi}|^2}\Big\}}.
\end{split}
\end{equation}

Therefore, we can transform the original minimizing RMSE into maximizing joint positioning similarity coefficient to achieve user positioning, which can be modeled as
\begin{equation}
\setcounter{equation}{10}
\begin{aligned}
\max_{\{\hat{\mathbf{\Theta}}_{k},\hat{\mathbf{\Psi}}_{k}:\forall k\}} \quad &\sum_{k \in \mathcal{K}} \mathbf{\Xi}(\mathbf{\Theta}_k,\mathbf{\Psi}_k,\hat{\mathbf{\Theta}}_{k},\hat{\mathbf{\Psi}}_{k}),\\
\quad \mbox{s.t.} \qquad \,\, & \, \hat{\mathbf{\Theta}}_{k} > \mathbf{0},\hat{\mathbf{\Psi}}_{k} > \mathbf{0}, k \in \mathcal{K},
\label{eq11}
\end{aligned}
\end{equation}
where $\mathbf{\Theta}_k$ and $\mathbf{\Psi}_k$ are known, $k \in \mathcal{K}$.

Obviously, we can notice that equation (10) is non-convex, and the computational and spatial complexity of conventional user positioning schemes is prohibitively high,
making them incompatible with cell-free mMIMO systems. Therefore, in the following section, we introduce a novel two-stage distributed collaborative user positioning architecture with MARL to overcome the aforementioned challenges.
\section{MARL-based User Positioning Architecture}
In this section, we propose a two-stage user positioning architecture with MARL, which includes a preliminary positioning network based on received RSS values and an auxiliary correction network based on AOA values, termed distributed collaborative positioning (DCP)-multi-agent deep deterministic policy gradient (MADDPG) algorithm.
\subsection{Markov Decision Process Model}
Recently, many effective MARL algorithms have been derived, including MADDPG, where all agents obtain feedback through continuous interaction with the environment to collaborate in formulating effective strategies. Particularly, we can describe the proposed user positioning architecture with a MARL environment $<\mathcal{S}, \mathcal{A}, \mathcal{P}, \mathcal{R}, \gamma>$, where $\mathcal{S}$ is the observation space for all agents, $\mathcal{A}$ is the action space, $\mathcal{R}$ denotes the expected reward function, $\mathcal{P}:(\mathcal{S},\mathcal{A})\rightarrow\mathcal{S}$ is the state transition function, and $\gamma$ denotes the discounted factor.
%\addtolength{\topmargin}{-0.02in}
\subsection{Distributed Collaborative Positioning Network}
In our MARL-based user positioning architecture, all APs are considered as entities that directly interact with cell-free mMIMO systems, and all antennas deployed by the same AP are considered as a whole for analysis. Then, we consider a novel MADDPG-based distributed collaborative positioning network to achieve user positioning, which consists of a RSS-based preliminary positioning network and an AOA-based supplementary correction network, as shown in Fig. 1.
\subsubsection{RSS-based Preliminary Positioning Network}
Each AP maps its observed RSS value $\boldsymbol{s}_{t,m}^{{r}} = [{\psi}_{m1},\ldots,{\psi}_{mK}]^{T}$ to estimated values $\boldsymbol{a}_{t,m}^{r} = [\hat{d}_{m1},\ldots,\hat{d}_{mK},\hat{\theta}_{m1}^{r},\ldots,\hat{\theta}_{mK}^{r}]^{T}$, $m \in \mathcal{M}$. Then, we can adopt the normalized distance-based positioning similarity coefficient $\mathbf{\bar{\Xi}}^{\mathrm{r}}(\mathbf{\Psi}_k,\hat{\mathbf{\Psi}}_{k})$ to quantify the reward of this preliminary positioning network $\boldsymbol{r}_{t}^{r} = [r_{t,1}^{r},\ldots,r_{t,M}^{r}]$ with $r_{t,m}^{r} = \sum_{k \in \mathcal{M}} \mathbf{\bar{\Xi}}^{\mathrm{r}}(\mathbf{\Psi}_k,\hat{\mathbf{\Psi}}_{k,m})$, where $\hat{\mathbf{\Psi}}_{k,m}$ is the RSS value extracted from the position of UE $k$ estimated by AP $m$. Correspondingly, the policy gradient of this network for $\pi_{m}^{r}$ can be modeled as
\begin{equation}
\setcounter{equation}{11}
\begin{aligned}
\nabla_{\theta_{\pi_m^r}}J(\theta_{\pi_m^{r}})=
{\mathbb{E}
\Big[\nabla_{\theta_{\pi_m^r}}\pi_m^r(\boldsymbol{a}_{t,m}^{r})
Q_{\theta_{Q_{\pi_m^r}}}(\boldsymbol{s}_{t}^{r},\boldsymbol{a}_{t}^{r})\Big]}.
\label{eq12}
\end{aligned}
\end{equation}

Besides, it is obvious that the action value $Q_{\theta_{Q_{\pi_m^r}}}(\boldsymbol{s}_{t}^{r},\boldsymbol{a}_{t}^{r})$ is calculated by the current critic network ${{\theta_{Q_{\pi_m^r}}}}$, then the mean-squared Bellman error function can be expressed as \cite{[5]}
\begin{equation}
\setcounter{equation}{12}
\begin{split}
L({{\theta_{Q_{\pi_m^r}}}}) = \mathbb{E}_{\boldsymbol{s}_{t}^{r},\boldsymbol{a}_{t}^{r}\sim \mathcal{D}^{r}}\Big[\Big(Q_{{\theta_{Q_{\pi_m^r}}}}(\boldsymbol{s}_{t}^{r},\boldsymbol{a}_{t}^{r})-y_{t,m}^{r}\Big)^2\Big],
\label{eq13}
\end{split}
\end{equation}
where $\mathcal{D}^{r}$ is the replay buffer of the positioning network, $y_{t,m}^{r}=r_{t,m}^{r}+\gamma^{r}(Q_{{\theta_{Q_{\pi_m^{p,o}}}}}(\boldsymbol{s}_{t}^{p,o},\boldsymbol{a}_{t}^{p,o}))$ is the target value with the target critic value $Q_{{\theta_{Q_{\pi_m^{p,o}}}}}(\boldsymbol{s}_{t}^{p,o},\boldsymbol{a}_{t}^{p,o})$.

Furthermore, we can adopt the soft update $\tau^r  \ll 1$ to ensure the stability of the target actor network and critic network.
\subsubsection{AOA-based Supplementary Correction}
By contrast, designing a correction network using AOA values can compensate for the neglect of angle information in the preliminary positioning network.
Then, we define the state and action as $\boldsymbol{s}_{t,m}^{a} = [{\hat{\theta}}_{m1}^{r},\ldots,{\hat{\theta}}_{mK}^{r}]^{T}$ and $\boldsymbol{a}_{t,m}^{a} = [{\Delta{\hat{\theta}}_{m1}^{a}},\ldots,{\Delta{\hat{\theta}}_{mK}^{a}}]^{T}$, respectively. Correspondingly, we adopt the joint similarity coefficient to quantify the reward $\boldsymbol{r}_{t}^{a}$ with $r_{t,m}^{\mathrm{a}}=\sum_{k \in \mathcal{K}}$ $\mathbf{\Xi}(\mathbf{\Theta}_k,\mathbf{\Psi}_k,\hat{\mathbf{\Theta}}_{k,m},\hat{\mathbf{\Psi}}_{k,m})$, where $\hat{\mathbf{\Theta}}_{k,m}$ is the AOA value extracted from the position of UE $k$ estimated by AP $m$.

Similarly, the policy gradient for $\pi_{m}^{a}$ can be modeled as
\begin{equation}
\setcounter{equation}{13}
\begin{aligned}
\nabla_{\theta_{\pi_m^a}}J(\theta_{\pi_m^{a}})={\mathbb{E}\Big[\nabla_{\theta_{\pi_m^a}}\pi_m^a(\boldsymbol{a}_{t,m}^{a})
Q_{\theta_{Q_{\pi_m^a}}}(\boldsymbol{s}_{t}^{a},\boldsymbol{a}_{t}^{a})\Big]}.
\label{eq14}
\end{aligned}
\end{equation}

Additionally, the mean-squared Bellman error function of the current critic network ${\theta_{Q_{\pi_m^a}}}$ can be defined as
\begin{equation}
\setcounter{equation}{14}
\begin{split}
L({{\theta_{Q_{\pi_m^a}}}}) = \mathbb{E}_{\boldsymbol{s}_{t}^{a},\boldsymbol{a}_{t}^{a}\sim \mathcal{D}^{a}}\Big[\Big(Q_{{\theta_{Q_{\pi_m^a}}}}(\boldsymbol{s}_{t}^{a},\boldsymbol{a}_{t}^{a})-y_{t,m}^{\mathrm{c}}\Big)^2\Big],
\label{eq15}
\end{split}
\end{equation}
where $\mathcal{D}^{a}$ is the replay buffer, the action value $Q_{\theta_{Q_{\pi_m^a}}}(\boldsymbol{s}_{t}^{a},\boldsymbol{a}_{t}^{a})$ is calculated by the current critic network ${{\theta_{Q_{\pi_m^\mathrm{c}}}}}$, and $y_{t,m}^{a}=r_{t,m}^{a}+\gamma^{a}(Q_{{\theta_{Q_{\pi_m^{a,o}}}}}(\boldsymbol{s}_{t}^{a,o},\boldsymbol{a}_{t}^{a,o}))$ is the target value with the target critic value $Q_{{\theta_{Q_{\pi_m^{a,o}}}}}(\boldsymbol{s}_{t}^{a,o},\boldsymbol{a}_{t}^{a,o})$.
Moreover, the soft update is carried out with $\tau^a \ll 1$ to ensure that the target actor network and critic network remains stable.
\addtolength{\topmargin}{-0.02cm}
\section{Numerical Results}
In this section, a cell-free mMIMO system is investigated, where $M$ APs and $K$ UEs are uniformly distributed in a $100 \times 100\,\mathrm{m^2}$ area with a wrap-around scheme to evaluate the positioning accuracy of the proposed DCP-MADDPG scheme \cite{[6]}. Then, the height difference, the carrier frequency, and the coherence bandwidth are modeled as 10 m, 10 GHz, and 200 kHz, respectively. Furthermore, for the baseline scheme composed of fingerprint positioning, we selected the corresponding reference point spacing $\eta$ as 0.5-2.5 m to better analyze the positioning accuracy and computational complexity.
\begin{figure}[t]
	\centering
    \includegraphics[scale=0.45]{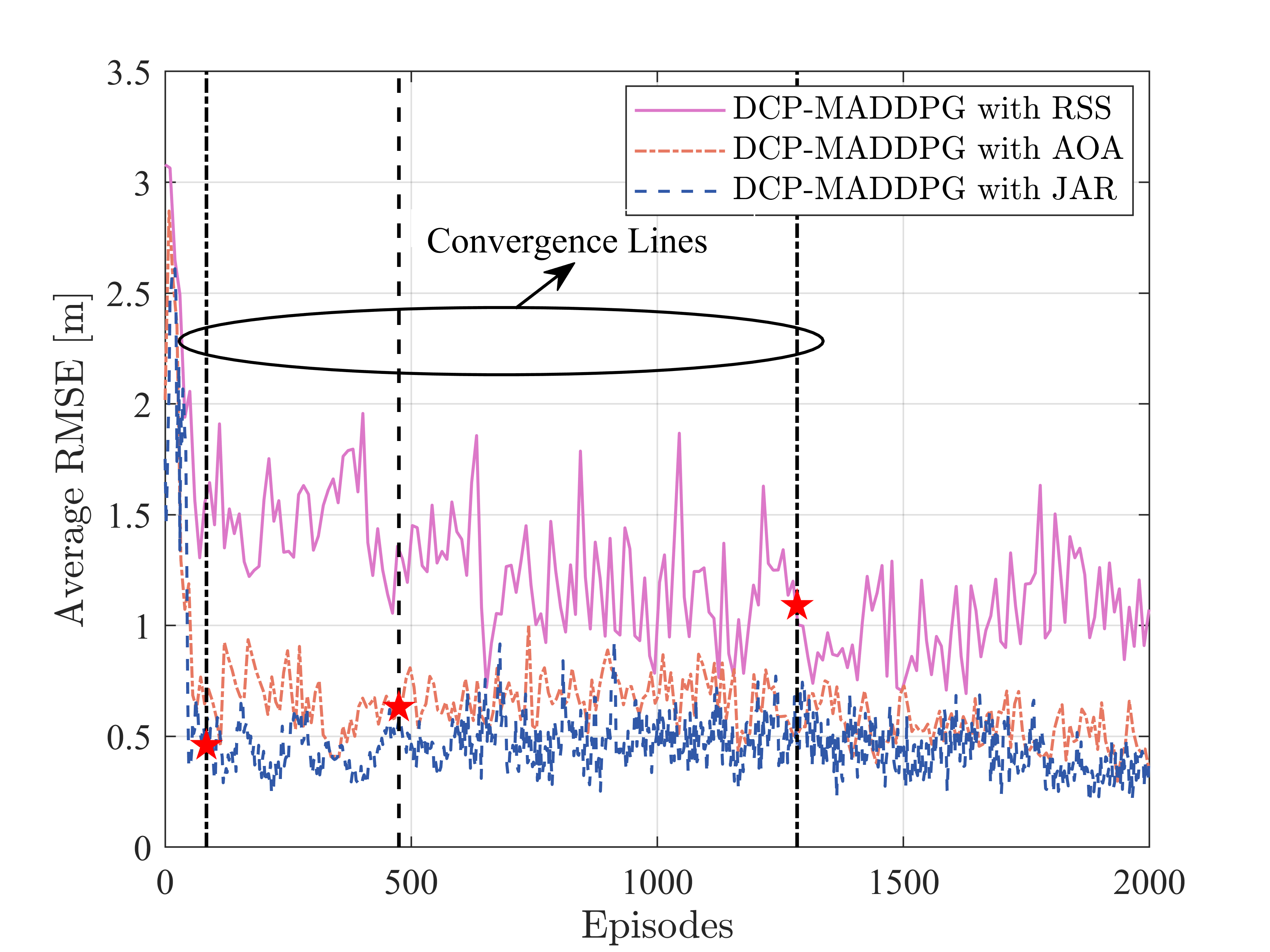}
    \caption{Convergence rate over different MARL-based positioning schemes with $M=36$, $K=\tau_p=6$, $L=8$, and $\Delta = \lambda/2$.}
\end{figure}
\begin{figure}[t]
	\centering
    \includegraphics[scale=0.45]{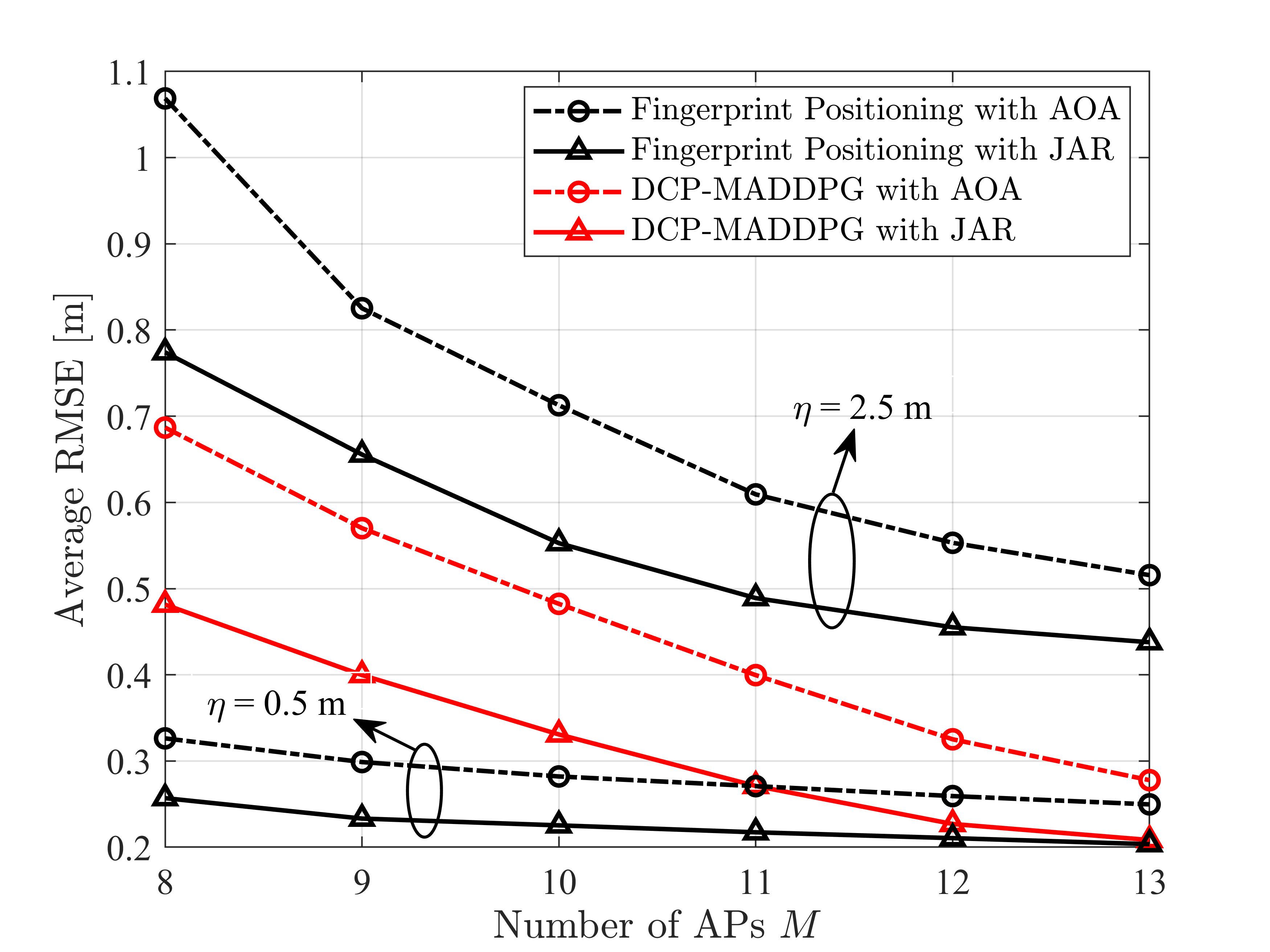}
    \caption{The average RMSE versus the number of APs with $K=\tau_p=6$, $L=8$, and $\Delta = \lambda/2$.}
\end{figure}
\begin{figure}[t]
	\centering
    \includegraphics[scale=0.45]{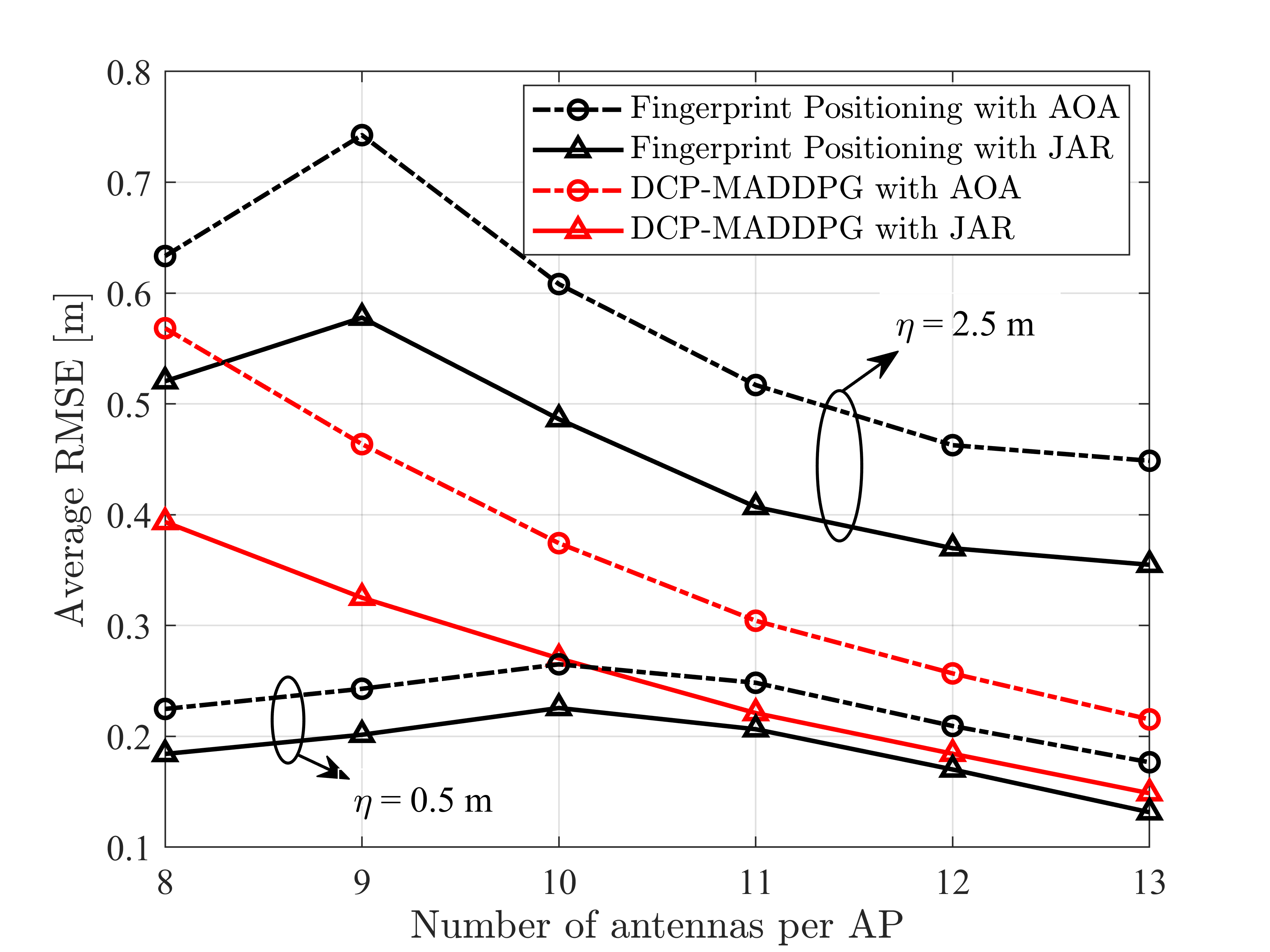}
    \caption{The average RMSE versus the number of antennas per AP with $M=36$, $K=\tau_p=6$, and $\Delta = \lambda/2$.}
\end{figure}

We firstly investigate the effect of adopting different positioning information as the state space for MARL networks. Fig. 2 shows the convergence rate over various MARL-based positioning schemes with different positioning information, including RSS, AOA, and joint AOA-and-RSS (JAR). Compared with conventional MARL-based schemes that utilizes a single positioning information, e.g., RSS-based or AOA-based information, the proposed JAR-based positioning scheme achieves positioning by fully combining distance and angle information of all UEs extracted from channel state information, thereby improving convergence rates by 92.33\% and 79.32\%, respectively. This demonstrates the effectiveness of reasonably combining distance and angle information in enhancing the convergence rate of the DPC-MADDPG algorithm.

Moreover, Fig. 3 and Fig. 4 investigate the average RMSE as a function of the number of APs $M$ and antennas per AP $N$, respectively.
Note that increasing the number of reference points has a significant impact on reducing estimation errors under fingerprint positioning algorithms. This reduction occurs as the reference point spacing decreases, resulting in a positioning performance improvement of over 51.58\% when the comparison reference point spacing $\eta$ decreases from 2.5 m to 0.5 m. Specifically, Fig. 3 indicates that as the number of APs increases, the positioning error gradually decreases. Our proposed DCP-MADDPG scheme's positioning performance gradually approaches that of fingerprint positioning with $\eta = 0.5$, indicating that increasing the number of APs facilitates their collaboration, thereby obtaining more accurate angle and distance information to enhance positioning.

On the other hand, Fig. 4 demonstrates that compared to conventional fingerprint positioning schemes, our proposed DCP-MADDPG scheme achieves similar positioning results as Fig. 3, which can better balance positioning performance and computational complexity. This is because the proposed scheme adopts a two-stage joint network composed of a RSS-based preliminary positioning network and an AOA-based auxiliary correction network, which optimizes positioning performance through mutual cooperation.

\section{Conclusion}
In this paper, we derived RSS values and the angular domain power matrix for user positioning model with DFT matrix in cell-free mMIMO systems. To address the challenges of high computational complexity and poor scalability in conventional positioning schemes, we proposed a novel distributed collaborative positioning scheme with MARL, which includes a two-stage structure consisting of a preliminary positioning network and an auxiliary correction network.
Finally, numerical results verified that the proposed scheme can effectively achieve a balance between positioning accuracy and computational complexity. In future research, we aim to expand our investigation from the far-field region to the near-field region, exploring user positioning challenges in cross-field environments.
\bibliographystyle{IEEEtran}
\bibliography{IEEEabrv,Ref}
\end{document}